\documentclass[aps,nofootinbib,preprint,showpacs]{revtex4}
\usepackage{array}
\usepackage{amsmath}
\usepackage{graphicx}
\usepackage{amssymb}

\begin{document}
\preprint{ANL-HEP-PR-09-93, EFI-09-24} 
\title{Next-to-Leading Order Cross Sections for New Heavy Fermion Production
at Hadron Colliders}
\author{\vspace{0.5cm} Edmond L. Berger}
\email{berger@anl.gov}
\affiliation{High Energy Physics Division, Argonne National Laboratory, Argonne,
Illinois 60439, USA}
\author{Qing-Hong Cao}
\email{caoq@hep.anl.gov}
\affiliation{High Energy Physics Division, Argonne National Laboratory, Argonne,
Illinois 60439, USA}
\affiliation{Enrico Fermi Institute, University of Chicago, Chicago, Illinois 60637, USA}

\begin{abstract}
We evaluate the cross sections for new heavy fermion production at
three Large Hadron Collider energies accurate to next-to-leading order
in perturbative quantum chromodynamics. We treat the cases of pair
production of heavy quarks via strong interactions, single heavy quark
production via electroweak interactions, and the production of heavy
leptons. Theoretical uncertainties associated with the choice of the
renormalization scale and the parton distribution functions are specified.
We derive a simple and useful parametrization of our results which
should facilitate phenomenological studies of new physics models that
predict new heavy quarks and/or leptons.
\end{abstract}

\pacs{12.38.Bx, 12.38.Qk, 13.85.Ni, 14.65.Jk}

\maketitle

%*********************************************
%
%  Introduction 
%
%*********************************************
\section{Introduction} 
Although the standard model (SM) of particle
physics consists of three fermion generations, the number of these
generations is not fixed by theory. Asymptotic freedom in quantum
chromodynamics (QCD) limits the number of generations to fewer than
nine. Neutrino counting, based on data at the intermediate vector
boson $Z$, shows that the number of generations having light neutrinos
($m_{\nu}\ll m_{Z}/2$ ) is equal to 3. However, neutrino oscillations
suggest a new mass scale that is beyond that of the SM, and the possibility
of additional heavier neutrinos is open. In addition, it has recently
been emphasized that the electroweak oblique parameters do not exclude
a fourth generation of chiral fermions~\cite{Alwall:2006bx,Kribs:2007nz,
Soni:2008bc,Bobrowski:2009ng,Chanowitz:2009mz,Holdom:2009rf,Soni:2009fg,
Eilam:2009hz}.
In the era of the Large Hadron Collider (LHC), the search for heavy
quarks and leptons beyond those of the SM should be kept in mind.

\begin{figure}[b]
\includegraphics[scale=0.4]{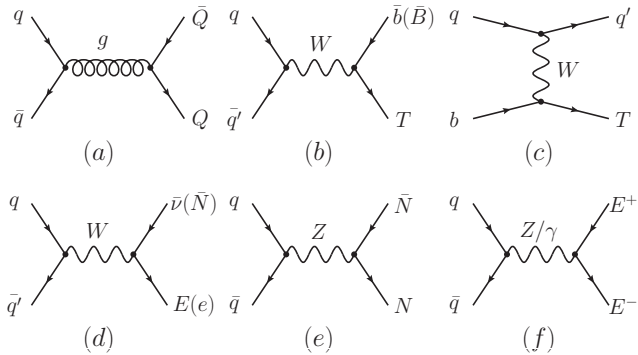}
\caption{Representative leading order Feynman diagrams for new heavy fermion
production. Uppercase (lowercase) letters denote new heavy (SM) fermions.
The usual gauge bosons are denoted by $g,W,Z,\gamma$. \label{fig:feyn-diag}}
\end{figure}

Observation of new heavy fermions requires knowledge of their expected
production cross sections and decay properties. In this work, we present
inclusive cross sections of fourth generation fermion production at
three LHC energies, either in pairs or singly, calculated to next-to-leading
(NLO) order accuracy in QCD. Leading order Feynman diagrams for new
heavy fermion production are shown for illustrative purposes in Fig.~\ref{fig:feyn-diag}.
We do not show the full set of NLO diagrams used in our calculations.
Theoretical uncertainties associated with higher-order perturbative
contributions and with the choice of parton distribution functions
(PDFs) are specified. We show that the cross sections can be fitted
with a rather simple analytic formula, and we present the functional
form and values of the parameters for the heavy fermion production
cross sections, and their dependence on the mass of the heavy fermion.

%*********************************************
%
%  Heavy quark pair production in QCD 
%
%*********************************************
\section{Heavy quark pair production in QCD} 
Heavy quarks that carry
the color charge exist in many models of new physics (NP). Examples
include models with a sequential fourth generation~\cite{Amsler:2008zzb,Kribs:2007nz},
or with a vectorlike $t$ prime ($T$)~\cite{delAguila:2000rc,Dobrescu:2009vz},
or with a $b$ prime ($B$)~\cite{Choudhury:2001hs}, or with a heavy
top quark partner~\cite{ArkaniHamed:2002qy,Perelstein:2003wd,Han:2003wu,Cheng:2005as}.
According to their physics motivation, the NP models can be further
categorized into two classes. In the first group, heavy quarks are
introduced to explain discrepancies between data and SM predictions.
In the second, heavy quarks are essential to solve theoretical problems.
For example, in Little Higgs models~\cite{ArkaniHamed:2002qy}, the
heavy top partner cancels the quadratic divergence in the quantum
corrections to the Higgs boson mass associated with the SM top quark
loop. The phenomenology of heavy quark identification at the LHC has 
been studied extensively~\cite{delAguila:1989rq,AguilarSaavedra:2005pv,
Matsumoto:2006ws,Freitas:2006vy,Meade:2006dw,Cao:2006wk,Belyaev:2006jh,
Contino:2008hi,AguilarSaavedra:2009es}. 

\begin{figure}
\includegraphics[scale=0.47]{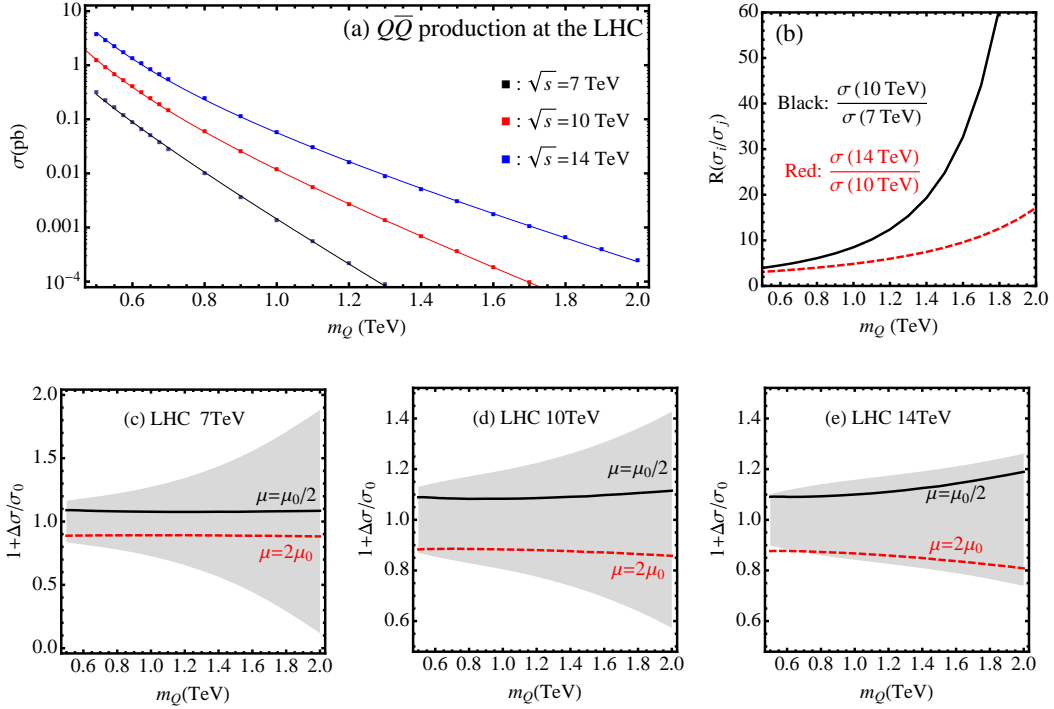}
\caption{(a) NLO $Q\overline{Q}$ pair production cross sections as a function of the quark
mass $m_{Q}$ at the LHC.  The square symbols denote the results of
the exact calculation, while the curves present results of our phenomenological
fit. (b) Ratios of the cross sections at different c.m. energies.
(c-e) Theoretical uncertainties of the $Q\overline{Q}$ production cross
section. The band denotes the PDF uncertainties, while the black solid
(red dashed) curve denotes the scale dependence obtained by varying
the renormalization scale by a factor 2 about the central value $\mu_{0}=m_{Q}$.
\label{fig:QQ}}

\end{figure}

Similar to the top quark in the SM, heavy quark $Q$ can be pair
produced via the strong interaction processes, 
\[
q\bar{q}\to g\to Q\overline{Q}\quad{\rm and}\quad gg\to Q\overline{Q},\]
where $g$ denotes a gluon. In this work we assume the coupling $g$-$Q$-$\bar{Q}$
is the same as the gluon-quark interaction in the SM. The cross section
for a different coupling strength can be obtained easily from our
results by rescaling. In Fig.~\ref{fig:QQ}(a) we plot our calculated
NLO inclusive cross sections for $Q\overline{Q}$ production at the LHC
for three choices of the proton-proton center mass (c.m.) 
energy.~\footnote{Our numerical code for the NLO $Q\overline{Q}$ inclusive 
cross section is based on the analytic results in 
Refs.~\cite{Nason:1987xz,Beenakker:1988bq,Nason:1989zy,Beenakker:1990maa}.}
The square symbol denotes the results of our exact NLO calculation
whereas the curve is drawn from the simple fitting functions discussed
below. As shown by the ratios presented in Fig.~\ref{fig:QQ}(b),
an increase in the c.m. energy from $7$ to $14\,{\rm TeV}$
can enhance the total cross section markedly, depending on the value
of $m_{Q}$.  The values of the standard model parameters used in our 
calculations are found in Appendix A.  

There are uncertainties in the predicted NLO cross sections associated with 
the choices of the renormalization scale ($\mu_{R}$) and the factorization 
scale ($\mu_{F}$) and with the choice of the PDFs.  
We elect to use the CTEQ6.6M PDFs~\cite{Stump:2003yu} for 
our evaluations.   Other sets of PDFs are available, notably the 
MSTW~\cite{Martin:2009iq} and NNPDF~\cite{Ball:2008by} sets, 
and one might use these sets in addition to the CTEQ6.6M set to 
compute a range of predictions.  All of these PDF sets include  
an analysis of the uncertainties in the PDF determinations associated 
both with uncertainties in the data used in the global fits and with the 
choices of theoretical expressions used to fit the data.  A comparison 
may be found in Ref.~\cite{Martin:2009iq}.   In this paper, we compute 
the PDF uncertainty (at $90\%$ C.L.) from the master formula in Eq.~(2.5) in 
Ref.~\cite{Stump:2003yu}, using all 44 sets of the CTEQ6.6M package. 
For heavy quark production at the large masses we are considering, 
the PDF uncertainties arise primarily from the valence quark PDF 
where different PDF sets tend to agree better than they do for gluon 
distributions. 

In Fig.~\ref{fig:QQ}(c)-\ref{fig:QQ}(e) 
we display the uncertainties associated
with the choices of PDFs and the renormalization/factorization
scale. 
~\footnote{
For the scale dependence, $\Delta \sigma / \sigma_0$ is defined as
$(\sigma(\mu_i)-\sigma(\mu_0))/\sigma(\mu_0)$, while for the PDF dependence 
$\Delta \sigma$ is defined as
$\Delta \sigma= 1/4 \sum_i^n ( \sigma_i^{+}-\sigma_i^{-})^2$, where  
where $\sigma_i^{\pm}$ denotes the upper (lower) cross section
for the $i$th eigenvector of the PDF sets and $n=22$ for CTEQ6.6M PDF sets.
The PDF uncertainties are then plotted as
$1 \pm \Delta \sigma / \sigma_0$ where $\sigma_0$ stands for the cross section
of the averaged PDF set. This definition applies to all other figures with 
cross section uncertainties.}
The uncertainties are portrayed relative to the cross section with the 
best-fit PDF by the bands in Fig.~\ref{fig:QQ}. The PDF uncertainties are
relatively large because the mass of the heavy quark requires that  
the PDFs be sampled in regions where they are relatively unconstrained.
The PDF uncertainties decrease with increasing c.m. energy. 

The uncertainties in the NLO cross section associated with the renormalization
scale ($\mu_{R}$) and factorization scale ($\mu_{F}$) are shown
in Fig.~\ref{fig:QQ}. These uncertainties can be considered as an
estimate of the size of unknown higher-order contributions. In this
study, we set $\mu=\mu_{R}=\mu_{F}$ and vary it around the central
value of $\mu_{0}=m_{Q}$, where $m_{Q}$ is the mass of the heavy
quark. Typically, a factor of 2 is used as a rule of thumb, and we
display curves with $\mu=2\mu_{0}$ and $\mu=\mu_{0}/2$. 
In Figs.~\ref{fig:QQ}(c)-\ref{fig:QQ}(e)
we plot $1 + \sigma(\mu_{i})/\sigma(\mu_{0})$ as a function
of $m_{Q}$. The cross sections vary between about $-10\%$ for $\mu=2\mu_{0}$
and $+10\%\sim15\%$ for $\mu=\mu_{0}/2$. We note that the scale
dependence at the three energies is insensitive to $m_{Q}$. It is
comparable to the PDF uncertainties in the region of relatively small
$m_{Q}$ but is much smaller than the PDF uncertainties in the region
of large $m_{Q}$.

We find that our calculated cross sections can be fitted well by a
simple three parameter analytic expression: 
\begin{equation}
\log\left[\frac{\sigma(m_{Q},\mu)}{{\rm pb}}\right]
=A(\mu)\left(\frac{m_{Q}}{{\rm TeV}}\right)^{-1}
+B(\mu)+C(\mu)\left(\frac{m_{Q}}{{\rm TeV}}\right),
\label{eq:formula}
\end{equation}
where the units of $\sigma$ and $m_{Q}$ are picobarn (pb) and TeV.
The variation with respect to the reference cross section (corresponding
to $\mu=m_{Q}$) gives the NLO scale dependence $\Delta_{\mu}(m_{Q})$.
The values of $\Delta_{\mu}$, and coefficients $A$, $B$, $C$ are
listed in Table~\ref{tab:fitpara-QQ}. 
Our results are presented in the form
\begin{equation}
\sigma=\sigma(\mu_0)^{\Delta\sigma(\mu_0/2)}_{\Delta\sigma(2\mu_0)},
\end{equation}
where $\sigma(\mu_0)$ is our prediction at the scale $\mu_0$. The
cross sections at the scale $\mu_0/2$ and $2\mu_0$ read as 
$\sigma(\mu_0/2) = \sigma(\mu_0) + \Delta\sigma(\mu_0/2)$ and 
$\sigma(2\mu_0)  = \sigma(\mu_0) + \Delta\sigma(2\mu_0)$, respectively.

As is evident in Fig.~\ref{fig:QQ}(a),
the fitting function and parameters in Table~\ref{tab:fitpara-QQ}
provide an excellent representation of the calculated cross sections
over the range of heavy quark masses shown in the figure. We caution
that the expression should not be used outside this range. One could
extend the region of applicability of a fit to lower mass, say down
to the top quark mass, by simply increasing the number of terms in
the polynomial expression. 
In Appendix~\ref{sec:smallmass} we extend our fit of the cross sections for 
heavy quark pair
production to cover the wider mass range, 250 to 700~GeV.  We 
also present a tabulation showing our exact NLO calculation and our fitted
results.  The numerical comparison shows that our fitted values agree 
with the exact calculations well within $1\%$. 
Interested readers may contact the authors
for accurate fitting functions that cover the mass range from the
top quark mass to $4\,{\rm TeV}$.

\begin{figure}
\includegraphics[scale=0.47]{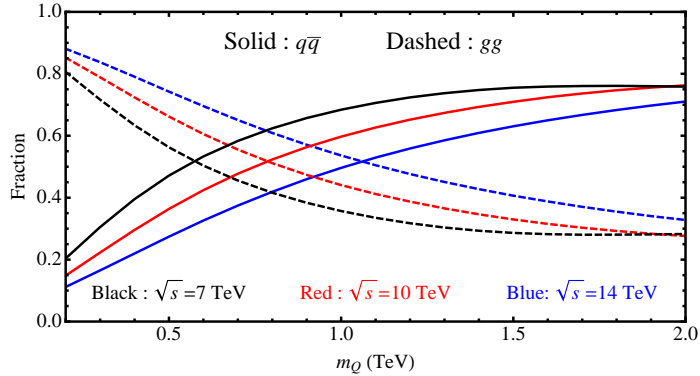}
\caption{Fractions of the $q\bar{q}$ and $gg$ initial state contributions
to the NLO total cross sections for $Q\overline{Q}$ pair production at
the LHC.\label{fig:fraction}}
\end{figure}

One reason more terms may be needed in a polynomial expression is
that processes with initial state gluons play an important role in
pair production of relatively light quarks. As shown in Fig.~\ref{fig:fraction},
at $\sqrt{s}=7\,{\rm TeV}$, the gluon initial state contribution
dominates over the quark initial state when $m_{Q}<550\,{\rm GeV}$;
at $\sqrt{s}=10\,{\rm TeV}$ when $m_{Q}<750\,{\rm GeV}$; while at
$\sqrt{s}=14\,{\rm TeV}$ when $m_{Q}<1\,{\rm TeV}$. For relatively
light quark production, say $m_{Q}<300\,{\rm GeV}$, $1/y^{2}$ and
$1/y^{3}$ terms ($y\equiv m_{Q}/{\rm TeV}$) are required to fit
the total cross section. 
In Appendix~\ref{sec:pdf}, we explore the apparent process
independence of the remarkably simple analytic expression Eq.~\ref{eq:formula}
and discuss its range of applicability in the heavy fermion mass.

\begin{table}
\caption{
Fitting parameters $(A,\, B,\, C)$ of the parametric formula
[Eq.~\ref{eq:formula}] for the NLO total cross sections (pb)
for heavy quark production at the LHC. 
The central values are given for the scale choice $\mu_{R}=\mu_{F}=\mu_0=m_{Q}$. 
The superscript and subscript denote the deviations in the coefficients 
from the central value when the scale is $\mu_{0}/2$  (superscript) 
or $2\mu_{0}$  (subscript).  We use the CTEQ6.6M PDFs. 
\label{tab:fitpara-QQ}}

\begin{tabular}{c|c|>{\centering}m{1.5in}>{\centering}m{1.5in}>{\centering}m{1.5in}}
\hline 
Process  & Parameter  & $\sqrt{s}=7\,{\rm TeV}$  & $\sqrt{s}=10\,{\rm TeV}$  & $\sqrt{s}=14\,{\rm TeV}$\tabularnewline
\hline
\hline 
 & $A$  & $\quad0.84569_{-0.03441}^{+0.04779}$  & $\quad1.76661_{-0.01912}^{+0.03079}$  & $\quad2.03833_{-0.01292}^{+0.02345}$\tabularnewline
$qq/gg\to Q\bar{Q}$  & $B$  & $\quad1.48655_{-0.02244}^{-0.05339}$  & $-0.39194_{-0.06751}^{+0.00546}$  & $-0.45930_{-0.08482}^{+0.03295}$\tabularnewline
 & $C$  & $-8.87048_{-0.08488}^{+0.09955}$  & $-5.82142_{-0.03726}^{+0.04314}$  & $-4.45853_{-0.01642}^{+0.01784}$\tabularnewline
\hline 
 & $A$  & $\quad0.72823_{-0.01262}^{+0.00130}$  & $\quad0.72901_{+0.00981}^{-0.00419}$  & $\quad0.69166_{+0.01911}^{-0.00600}$\tabularnewline
$qb\to q^{\prime}T$  & $B$  & $\quad1.66318_{+0.00630}^{-0.03563}$  & $\quad2.28354_{+0.00541}^{-0.01591}$  & $\quad2.94279_{-0.01098}^{-0.01395}$\tabularnewline
 & $C$  & $-3.38334_{-0.04100}^{+0.05251}$  & $-2.57503_{-0.03535}^{+0.03980}$  & $-2.06090_{-0.02618}^{+0.03918}$\tabularnewline
\hline 
 & $A$  & $\quad0.77742_{-0.00158}^{+0.00580}$  & $\quad0.78617_{+0.00520}^{+0.00133}$  & $\quad0.75829_{+0.00743}^{-0.00266}$\tabularnewline
$q\bar{b}\to q^{\prime}\bar{T}$  & $B$  & $\quad0.86229_{+0.03096}^{-0.04447}$  & $\quad1.53387_{+0.01617}^{-0.03343}$  & $\quad2.22983_{+0.01447}^{-0.01720}$\tabularnewline
 & $C$  & $-3.57353_{-0.05169}^{+0.05666}$  & $-2.72791_{-0.03900}^{+0.04759}$  & $-2.18287_{-0.03617}^{+0.03754}$\tabularnewline
\hline 
 & $A$  & $\quad1.37193_{+0.01703}^{-0.00541}$  & $\quad1.41641_{+0.00749}^{+0.00152}$  & $\quad1.46925_{+0.00596}^{-0.00453}$\tabularnewline
$q\bar{q^{\prime}}\to T\bar{b}$  & $B$  & $-3.44671_{-0.05685}^{+0.03710}$  & $-3.45805_{-0.02949}^{+0.01625}$  & $-3.33539_{-0.02316}^{+0.02351}$\tabularnewline
 & $C$  & $-5.22956_{-0.01643}^{+0.02415}$  & $-3.81301_{-0.01891}^{+0.02269}$  & $-3.00593_{-0.01352}^{+0.01228}$\tabularnewline
\hline 
 & $A$  & $\quad1.79403_{+0.02359}^{-0.00423}$  & $\quad1.77007_{+0.00967}^{-0.00699}$  & $\quad1.68940_{+0.00815}^{-0.00650}$\tabularnewline
$q\bar{q^{\prime}}\to\bar{T}b$  & $B$  & $-5.40590_{-0.06414}^{+0.02628}$  & $-5.03094_{-0.03479}^{+0.03140}$  & $-4.44604_{-0.02747}^{+0.02719}$\tabularnewline
 & $C$  & $-4.81694_{-0.01401}^{+0.02980}$  & $-3.67831_{-0.01460}^{+0.01489}$  & $-3.08803_{-0.01088}^{+0.00993}$\tabularnewline
\hline 
 & $A$  & $\quad0.09505_{+0.00202}^{-0.00087}$  & $\quad0.98804_{-0.00090}^{+0.01201}$  & $\quad1.07010_{+0.00304}^{-0.00263}$\tabularnewline
$q\bar{q^{\prime}}\to T\bar{B}$  & $B$  & $-1.08915_{-0.03419}^{+0.03084}$  & $-2.21256_{-0.00969}^{-0.01419}$  & $-2.20446_{-0.01452}^{+0.01778}$\tabularnewline
 & $C$  & $-10.4782_{-0.04719}^{+0.04744}$  & $-4.66219_{-0.03293}^{+0.04337}$  & $-3.75217_{-0.01956}^{+0.01638}$\tabularnewline
\hline 
 & $A$  & $\quad0.65738_{-0.00549}^{+0.01149}$  & $\quad1.15500_{+0.00280}^{-0.00182}$  & $\quad1.18663_{+0.00487}^{-0.00354}$\tabularnewline
$q\bar{q^{\prime}}\to\bar{T}B$  & $B$  & $-4.02175_{-0.00550}^{-0.01027}$  & $-5.09831_{-0.02683}^{+0.02561}$  & $-4.73348_{-0.02800}^{+0.02658}$\tabularnewline
 & $C$  & $-9.05880_{-0.06865}^{+0.07512}$  & $-6.08544_{-0.03100}^{+0.02944}$  & $-4.91903_{-0.01835}^{+0.01753}$\tabularnewline
\hline
\end{tabular}
\end{table}

In view of the interest in the top quark as the heaviest of the known
quarks, we provide NLO predictions of the top quark pair production
cross sections in Table~\ref{tab:SM-top} at three LHC energies for
a few values of $m_{t}$. Our result at $14\,{\rm TeV}$ is consistent
with Ref.~\cite{Kidonakis:2008mu}.

\begin{table}
\caption{NLO top quark pair production cross sections (pb) at the LHC 
with three different c.m. energies based on the CTEQ6.6M PDFs. 
The central values
are given for the scale choice $\mu_{R}=\mu_{F}=\mu_0=m_{t}$.  
The superscript and subscript denote the deviations of the coefficients 
from the central value when the scale is $\mu_{0}/2$  (superscript) or 
$2\mu_{0}$ (subscript). \label{tab:SM-top}
}
\begin{tabular}{c|ccc}
\hline 
$m_{t}({\rm GeV})$  & $\sqrt{s}=7\,{\rm TeV}$  & $\sqrt{s}=10\,{\rm TeV}$  & $\sqrt{s}=14\,{\rm TeV}$\tabularnewline
\hline
171  & $157.4_{-20.3}^{+18.9}$  & $396.5_{-48.4}^{+47.2}$  & $877.2_{-101.4}^{+103.1}$\tabularnewline
173  & $148.2_{-19.2}^{+17.7}$  & $374.5_{-45.8}^{+44.4}$  & $830.9_{-94.1}^{+93.9}$\tabularnewline
175  & $139.6_{-18.1}^{+16.5}$  & $354.0_{-43.4}^{+41.6}$  & $787.5_{-91.5}^{+91.3}$\tabularnewline
177  & $131.6_{-17.0}^{+15.6}$  & $335.0_{-41.1}^{+39.2}$  & $747.2_{-86.9}^{+86.4}$\tabularnewline
\hline
\end{tabular}
\end{table}

%*********************************************************
%
%  Heavy quark production through the EW interaction.
%
%********************************************************* 
\section{Heavy quark production through the electroweak interaction} 
Heavy quark production occurs also through electroweak interactions. One
example is the heavy top quark partner ($T$) in the Little Higgs
Model~\cite{Han:2003wu,Perelstein:2003wd}, which is responsible
for canceling large quantum corrections to the Higgs boson mass from
the SM top quark loop. In addition to its pair production through
QCD interactions, the $T$ quark can be produced singly in association
with a light quark through the processes \[
qb\to q^{\prime}T,\quad{\rm and}\quad q\bar{q^{\prime}}\to Tb.\]
It may also be produced in association with a heavy $B$ quark via
the $s$-channel subprocess 
\[ q\bar{q^{\prime}}\to W^{*}\to TB.\]

\begin{figure}
\includegraphics[scale=0.52]{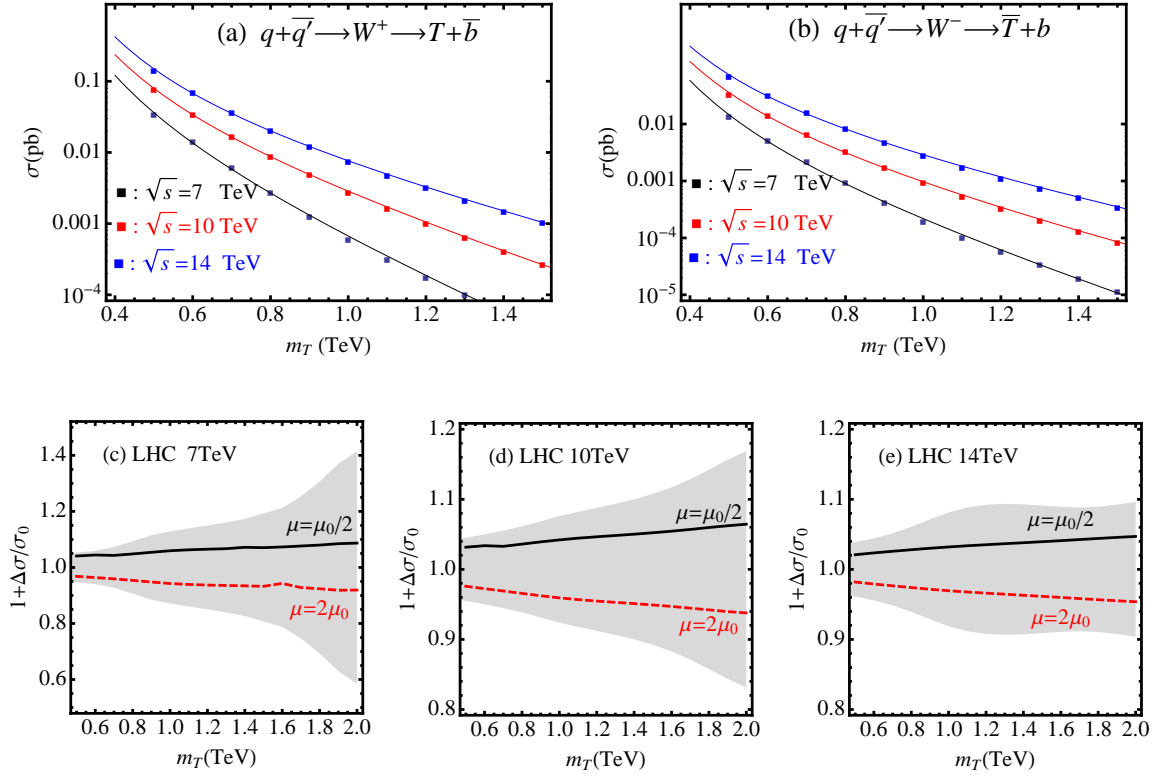}
\caption{NLO production cross sections for (a) $T\bar{b}$ and (b) $\bar{T}b$ 
at the LHC as a function of $m_T$, and (c-e) PDF and scale uncertainties of 
the $T\bar{b}$ production cross section. The $\overline{T}b$ production cross section 
exhibits similar uncertainties.   
\label{fig:xsec-Tb-schan} } 
\end{figure}

We consider first single $T$ quark production in association with
a SM $b$ quark via an $s$-channel subprocess, illustrated at leading order
in Fig.~\ref{fig:feyn-diag}(b). This subprocess involves a weak
interaction vertex $W$-$T$-$b$ and a quark mixing matrix $V_{Tb}$.
For simplicity, we assume that the coupling of $W$-$T$-$b$ is same
as the SM $W$-$t$-$b$ coupling and that $\left|V_{Tb}\right|=1$.
Our results may be rescaled for other choices. We calculate the $T\bar{b}$
and $\overline{T}b$ results separately since the $pp$ initial state at
the LHC is not a $CP$ eigenstate. The renormalization and factorization
scales are chosen to be $\mu_{R}=\mu_{F}=\sqrt{Q_{V}^{2}}$ where
$Q_{V}$ denotes the four-momentum of the $W$ boson in the propagator.

The square symbols in Fig.~\ref{fig:xsec-Tb-schan} show the predicted
production cross sections for $T\bar{b}$ and $\overline{T}b$ at the LHC
for three c.m. energies. The curves are our fit with the parametrization
in Eq.~\ref{eq:formula}. We note that this simple parametrization
fits single $T$ production via the $s$-channel process very well
for the three LHC energies. The fitting parameters are presented in
Table~\ref{tab:fitpara-QQ}. 
The cross section for single $T$ production is larger than the cross section 
for single $\overline{T}$ production by almost a factor $2\sim 3$.
This is due to the difference in parton densities of the colliding protons.
While in both cases the antiquark is from the quark sea of one of the incoming
protons, the probability that it collides with an up quark from the other
proton is higher than the probability for a collision with a down quark. 
Since $T\bar{b}$ and $\overline{T}b$
production exhibit similar uncertainties, we present only the results
for $T\bar{b}$ production in Figs.~\ref{fig:xsec-Tb-schan}(c-e).  
We note the scale dependence at the 
three energies is not very sensitive to $m_{T}$, and it is much
smaller than the PDF uncertainties in the large $m_{T}$ region.  An increase 
in the c.m. energy reduces the PDF uncertainties in the large $m_{T}$ region
sizably. 

\begin{figure}
\includegraphics[scale=0.5]{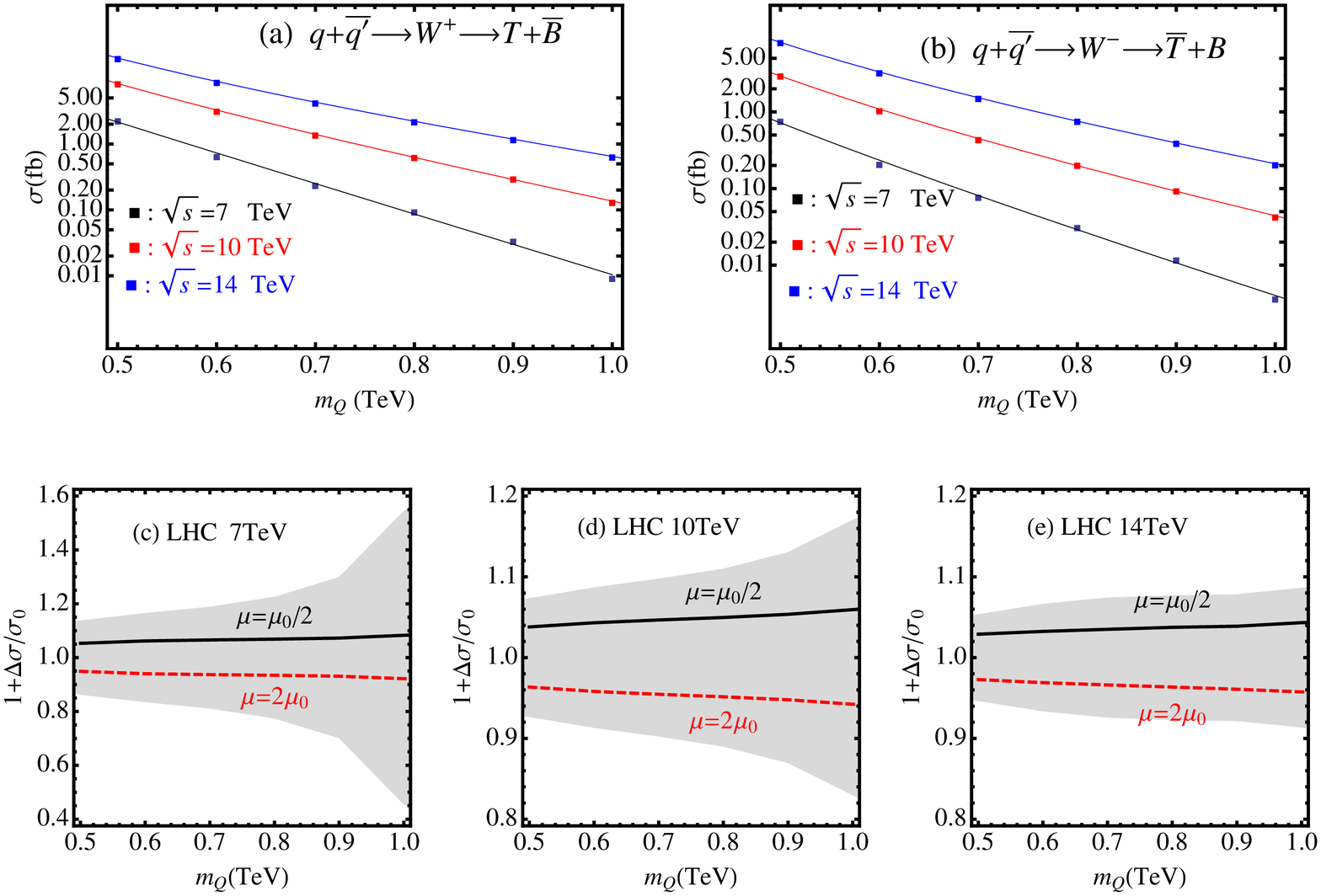}
\caption{NLO production cross sections for (a) $T\overline{B}$ and 
(b) $\overline{T}B$ at the LHC as a function of $m_Q = m_T = m_B$, 
and (c-e) PDF and scale uncertainties of the $T\overline{B}$ production cross section.
The $\overline{T}B$ production cross section exhibits similar uncertainties. 
\label{fig:xsec-TB-schan} }
\end{figure}

In the ``sequential'' fourth generation model, the heavy $T$
and $B$ quarks form an isodoublet, and their interaction with the
$W$ boson is identical to the $W$-$t$-$b$ interaction in the SM
with the substitution of $V_{tb}\to V_{TB}=1.$ For simplicity, we
take the $T$ and $B$ quarks to be degenerate, i.e. $m_{T}=m_{B}=m_{Q}$.
The $T$ and $B$ quarks can be produced together via the Drell-Yan
process (with an $s$-channel $W^{*}$). In Fig.~\ref{fig:xsec-TB-schan}
we show our calculated cross sections and the fitted curves based
on the parametrization in Eq.~\ref{eq:formula}. The parametrization
works well at $\sqrt{s}=10\,{\rm TeV}$ and $14\,{\rm TeV}$ but less
adequately at $7\,{\rm TeV}$. The inclusion of more terms in the
polynomial would improve the fit. The fitting parameters are listed
in Table~\ref{tab:fitpara-QQ}.

\begin{figure}
\includegraphics[scale=0.45]{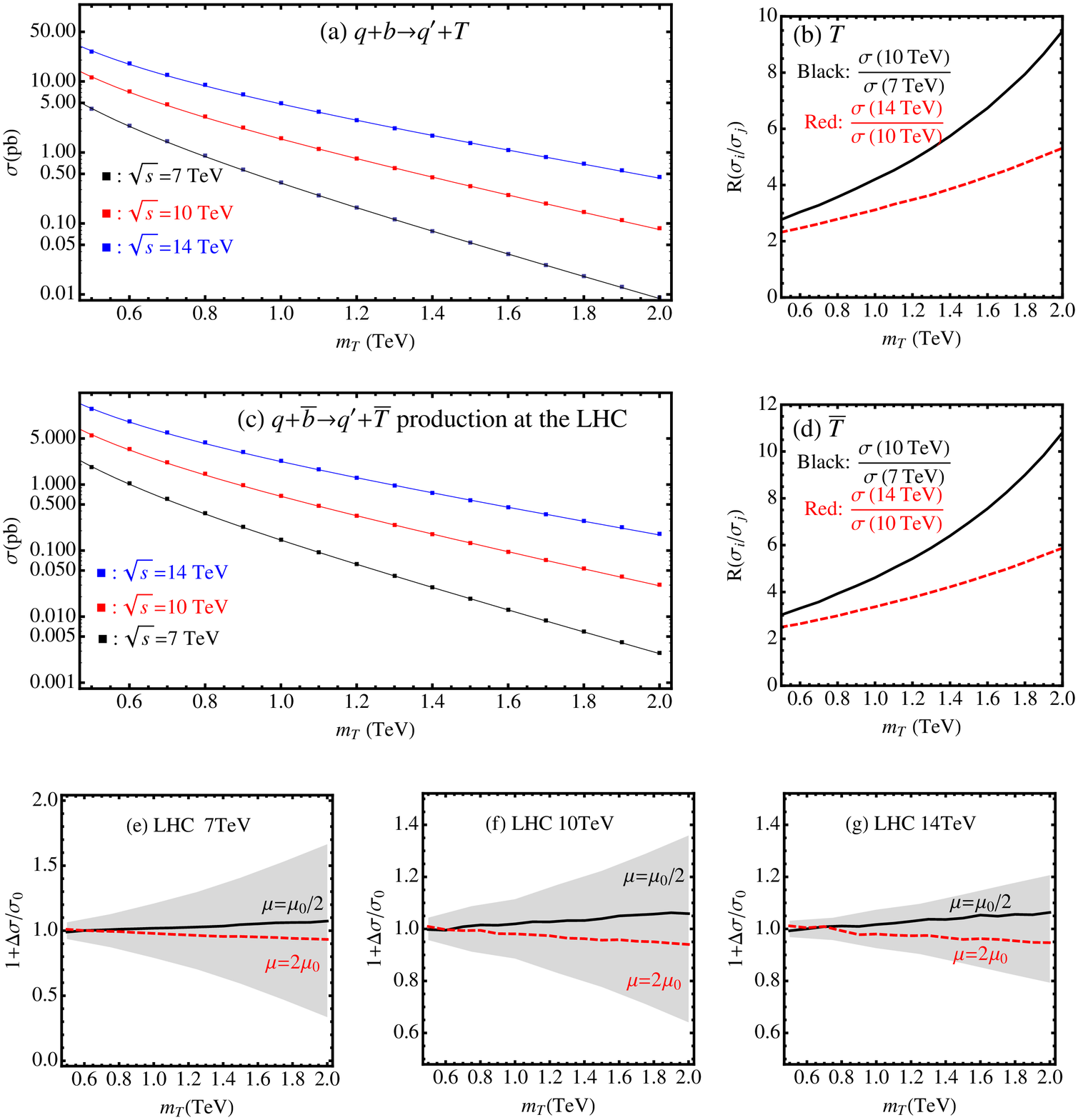}
\caption{Single heavy (a) $T$ production and (c) $\overline{T}$ production 
via the $t$-channel subprocess as a function of $m_T$.     
The square symbols denote the results of
the exact calculation while the curves present results of our phenomenological
fit. (b) and (d) Ratios of the cross sections at different c.m. energies.
(e-g) Theoretical uncertainties of the single $T$ production cross
section. The band denotes the PDF uncertainties while the black solid
(red dashed) curve denotes the scale dependence obtained by varying
the renormalization scale by a factor 2 about the central value $\mu_{0}=m_{T}$.
\label{fig:EWK_Q_tchan}}
\end{figure}

Last, we consider single $T$ quark production via a $t$-channel
exchange diagram, illustrated at leading order in Fig.~\ref{fig:feyn-diag}(c).
The production cross section is larger than in the $s$-channel
case because it is not subject to $1/s$ suppression. In this
calculation we adopt the so-called double deep inelastic-scattering
scale proposed in Ref.~\cite{Stelzer:1997ns}. Color conservation
enforces a natural factorization of the scales. The fermion line in
Fig.~\ref{fig:feyn-diag}(c) that does not include a heavy quark
probes a proton with the deep inelastic-scattering scale 
$\sqrt{Q^{2}_{V}}$ , which is identical
to the virtuality of the $W$ boson through NLO. The fermion line
that connects to a heavy quark sees the deep inelastic-scattering scale
for massive quarks of $\sqrt{Q^{2}_{V}+m_{Q}^{2}}$. 

Our simple parametrization also works well for the $t$-channel results
as may be seen in Figs.~\ref{fig:EWK_Q_tchan}(a) and 
\ref{fig:EWK_Q_tchan}(c). The fitting
parameters are found in Table~\ref{tab:fitpara-QQ}. Again, an increase
in the c.m. energy from $7$ to $14\,{\rm TeV}$ can enhance
the total cross section markedly, depending on the value of $m_{T}$;
see Figs.~\ref{fig:EWK_Q_tchan}(b) and \ref{fig:EWK_Q_tchan}(d). 
The PDF and scale uncertainties
are shown in Figs.~\ref{fig:EWK_Q_tchan}(e)-\ref{fig:EWK_Q_tchan}(g). 
Since single $T$ and single $\overline{T}$ production exhibit similar
uncertainties, we present only the results for single $T$ production.  
We note the scale dependence at the three energies is not sensitive to
$m_{T}$. The cross sections vary between $\pm2\%$ for $m_{T}\sim500\,{\rm GeV}$
to $\pm7\%$ for $m_{T}\sim2\,{\rm TeV}$.   The scale dependence is much smaller than
the PDF uncertainties in the large $m_{T}$ region.

%*****************************
%
%  Exotic lepton production.
%
%*****************************   
\section{Exotic lepton production} 
Several variants of the seesaw mechanism
have been proposed to explain light neutrino masses. Positive signals
could be observed at the LHC in the case that seesaw messengers exist
at the TeV scale or below. Three types of tree-level seesaw models
generate light neutrino Majorana masses via introduction of (1) a
right-handed neutrino singlet, (2) a complex scalar triplet $\Delta$
with hypercharge $Y=1$, and (3) a lepton triplet $\Sigma$ with $Y=0$.
In these seesaw models, neutrinos achieve their masses via a lepton
number violating operator at the scale $\Lambda$. The scale is not
necessarily very high, and it might be around a TeV. Searches for
these exotic leptons and, if observed, measurement of their properties
would verify and even distinguish different seesaw mechanisms~\cite{delAguila:2008cj}.
For this purpose, an accurate calculation including higher-order QCD
corrections is in order.

\begin{table}[b!]
\caption{Coupling strengths of the gauge vertices in heavy lepton pair production
for different models, where $g$ denotes the usual weak coupling strength.
Note that all the couplings are vectorlike.\label{tab:Coupling}}

\begin{tabular}{l>{\centering}m{0.8in}>{\centering}m{0.8in}>{\centering}m{0.8in}>{\centering}m{0.8in}}
\hline 
 & $WEN$  & $ZEE$  & $\gamma EE$  & $ZNN$\tabularnewline
\hline 
Majorana triplet  & $g$  & $g c_{W}$  & $e$  & $\cdots$\tabularnewline
Dirac triplet  & $g$  & $g c_{W}$  & $e$  & $\cdots$\tabularnewline
Lepton isodoublet  & ${\displaystyle \frac{g}{\sqrt{2}}}$  & ${\displaystyle \frac{g}{c_{W}}(-\frac{1}{2}+s_{W}^{2})}$  & $e$  & ${\displaystyle \frac{g}{2c_{W}}}$\tabularnewline
Charge singlet  & $\cdots$  & $\cdots$  & $e$  & $\cdots$\tabularnewline
\hline
\end{tabular}
\end{table}

We examine the following three possibilities for exotic lepton production:
\begin{eqnarray*}
 &  & q\bar{q}\to\gamma/Z\to E^{+}E^{-},\\
 &  & q\bar{q}\to Z\to NN,\\
 &  & q\bar{q^{\prime}}\to W\to EN,
\end{eqnarray*}
where $E(N)$ denotes the new heavy charged lepton (heavy neutrino).
The interactions of gauge bosons and heavy leptons are summarized
in Table~\ref{tab:Coupling} for four interesting NP models: (1)
a Majorana lepton triplet with $Y=0$,
(2) a Dirac lepton triplet with $Y=0$, (3) a lepton isodoublet, (4)
a charge singlet. The motivation for these models and more details
can be found in Ref.~\cite{AguilarSaavedra:2009ik}. It is worth 
mentioning the following properties: 
\begin{itemize}
\item All the interactions given in the Table~\ref{tab:Coupling} 
are vectorlike. 
\item There are no $ZNN$ interactions in the Majorana/Dirac lepton triplet
model because the triplet has zero hypercharge, and $NN$ pairs are
not produced. 
\item A Dirac lepton triplet is formed by two degenerate Majorana triplets
with opposite CP parities. As shown in Ref.~\cite{delAguila:2008hw},
the heavy lepton fields can be redefined in such a way that the Lagrangian
is written in terms of two charged leptons $E_{1}^{-}$, $E_{2}^{+}$
and a Dirac neutrino $N$. As a consequence of the presence of two
charged fermions instead of only one, the total heavy lepton production
cross section is twice that for a Majorana triplet. In this work,
however, we consider only one flavor of charged lepton in both the
Majorana triplet and the Dirac triplet. 
\end{itemize}
To make our results more useful, rather than focusing on a specific
NP model, we consider the following effective interaction in our numerical
calculation, 
\begin{align}
 & \mathcal{L}=-\frac{g}{c_{W}}\left(\overline{E}\gamma^{\mu}E
 + \overline{N}\gamma^{\mu}N\right)Z_{\mu}
 - \frac{g}{\sqrt{2}}\left(\overline{E}\gamma^{\mu}N W_{\mu}^{-}
 + \overline{N}\gamma^{\mu}E W_{\mu}^{+}\right)
 + e \overline{E}\gamma^{\mu}E A_{\mu},
\label{eq:coupling}
\end{align}
where the symbol $g$ stands for the usual weak coupling strength.
Numerical results for the models listed in Table~\ref{tab:Coupling}
can be easily derived from ours by simply rescaling the coupling strengths,
except for the process $q\bar{q}\to\gamma/Z\to E^{+}E^{-}$ where
there are interference effects of the photon and $Z$ boson contributions.
However, since the threshold for two heavy charged leptons is much
higher than the $Z$ boson mass, the interference effects are subleading.
We present separately the NLO cross sections for 
$q\bar{q}\to\gamma\to E^{+}E^{-}$ and $q\bar{q}\to Z\to E^{+}E^{-}$; 
the couplings involved in the two processes are overall factors. 
For all the $s$-channel processes listed above, the renormalization 
and factorization scales are chosen to be $\mu=\sqrt{Q_{V}^{2}}$ 
where $Q_{V}$ denotes the four-momentum of the gauge boson in the propagator.
For simplicity, we take the $E$ and $N$ leptons to be degenerate, i.e. 
$m_E = m_N = m_L$.

\begin{figure}
\includegraphics[scale=0.52]{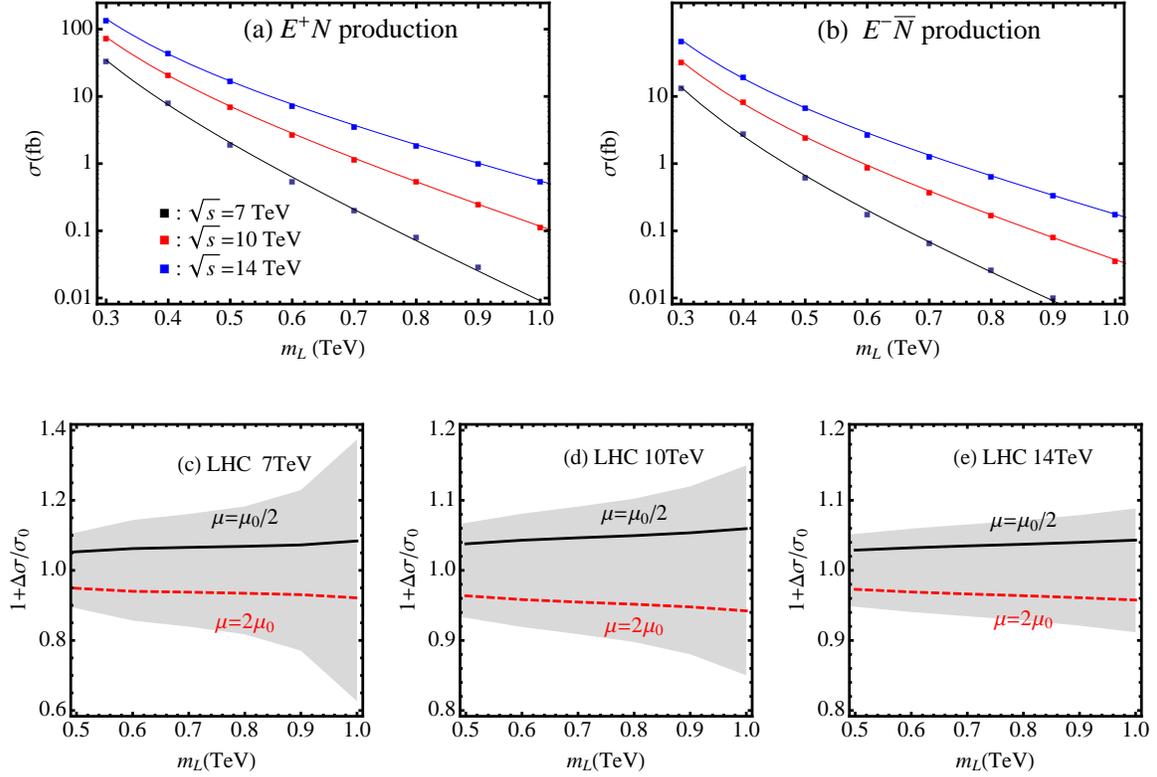}
\caption{NLO production cross sections for (a) 
$E^+ N$ and (b) $E^- \overline{N}$ production
at the LHC as a function of $m_L = m_E = m_N$, 
and (c-e) PDF and scale uncertainties of $E^+ N$ production.
The $E^- \overline{N}$ production cross section exhibits similar 
uncertainties.  
\label{fig:xsec-EN}} 
\end{figure}

In Figs.~\ref{fig:xsec-EN}(a) and \ref{fig:xsec-EN}(b) 
we display the NLO cross sections
for the $q\bar{q^{\prime}}\to W^{+}\to E^{+}N$ and 
$q\bar{q^{\prime}}\to W^{-}\to E^{-}\overline{N}$ processes, respectively.
Both processes exhibit similar scale and PDF
uncertainties as shown in Figs.~\ref{fig:xsec-EN}~(c)-\ref{fig:xsec-EN}(e). The fitting
parameters are given in Table~\ref{tab:fitpara-EN}. 

\begin{figure}
\includegraphics[scale=0.5]{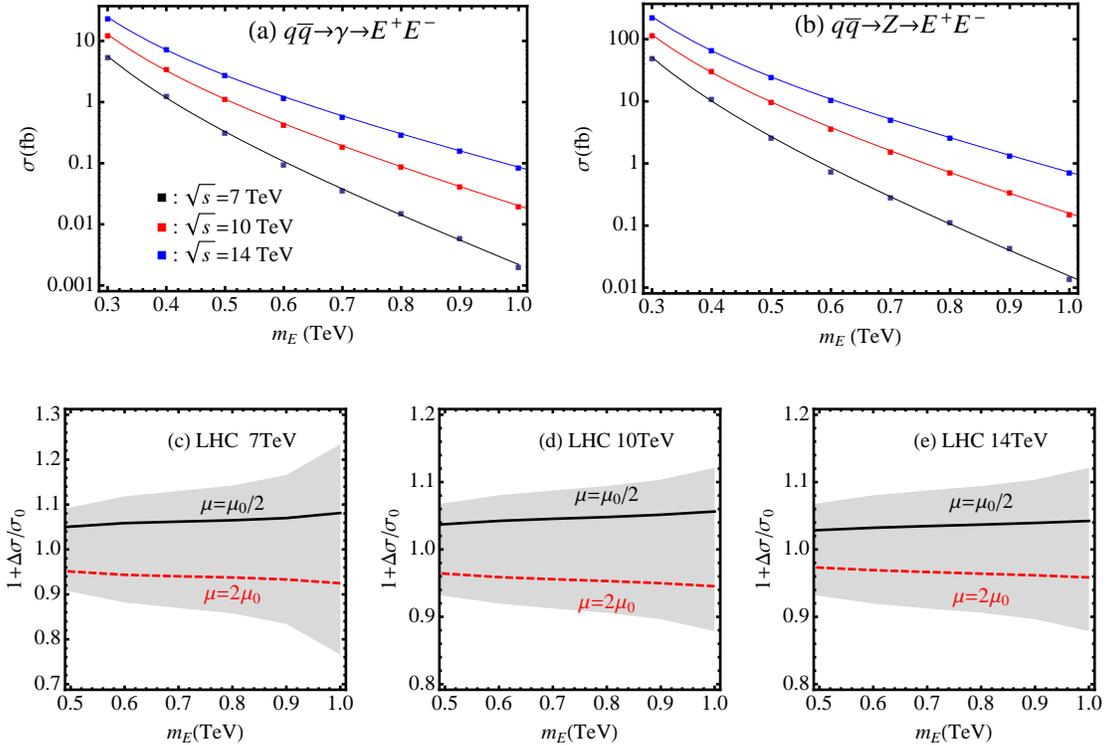}
\caption{NLO production cross sections for (a) 
$E^+ E^-$ production via an intermediate virtual photon and 
(b) via an intermediate $Z$ boson at the LHC as a function of $m_E$, 
and (c-e) PDF and scale uncertainties of $E^+ E^-$ production via
an intermediate virtual photon.  Production via an intermediate $Z$ boson 
exhibits similar uncertainties. 
\label{fig:xsec-EE}}
\end{figure}  

We also plot the NLO cross sections for the $q\bar{q}\to\gamma\to E^{+}E^{-}$
and the $q\bar{q}\to Z\to E^{+}E^{-}$ processes in Fig.~\ref{fig:xsec-EE}.
Both processes exhibit similar scale and PDF uncertainties as shown
in Fig.~\ref{fig:xsec-EE}(c)-\ref{fig:xsec-EE}(e). 
Note that the cross sections for
$q\bar{q}\to Z\to \overline{N}N$ are the same as those for 
$q\bar{q}\to Z\to E^{+}E^{-}$ owing to the same assignment of couplings
in Eq.~\ref{eq:coupling}; therefore, they are not shown here.

\begin{table}
\caption{
The fitting parameters $(A,\, B,\, C)$ in the parametric formula
[Eq.~\ref{eq:formula}] for the NLO total cross sections (pb) for heavy
lepton production at the LHC. The central values
are given for the scale choice $\mu_{R}=\mu_{F}=\mu_0=m_{E}$.  
The superscript and subscript denote the deviations of the coefficients 
from the central value when the scale is $\mu_{0}/2$  (superscript) 
or $2\mu_{0}$  (subscript). 
We use the CTEQ6.6M PDFs. 
\label{tab:fitpara-EN}}

\begin{tabular}{c|c|>{\centering}m{1.5in}>{\centering}m{1.5in}>{\centering}m{1.5in}}
\hline 
Process  & Parameter  & $\sqrt{s}=7\,{\rm TeV}$  & $\sqrt{s}=10\,{\rm TeV}$  & $\sqrt{s}=14\,{\rm TeV}$\tabularnewline
\hline
\hline 
 & $A$  & $\quad0.71010_{+0.00943}^{+0.00073}$  & $\quad0.77778_{+0.00406}^{-0.00278}$  & $\quad0.81919_{+0.00335}^{-0.00250}$\tabularnewline
$q\bar{q^{\prime}}\to E^{+}N$  & $B$  & $-2.90505_{-0.05227}^{+0.02141}$  & $-3.16717_{-0.02697}^{+0.02513}$  & $-3.11747_{-0.02179}^{+0.02228}$\tabularnewline
 & $C$  & $-9.41140_{-0.09655}^{+0.05659}$  & $-6.66323_{-0.03572}^{+0.03454}$  & $-5.20543_{-0.02455}^{+0.02221}$\tabularnewline
\hline 
 & $A$  & $\quad0.99258_{+0.00593}^{-0.00287}$  & $\quad0.98794_{+0.00560}^{-0.00431}$  & $\quad0.94199_{+0.00486}^{-0.00381}$\tabularnewline
$q\bar{q^{\prime}}\to E^{-}\bar{N}$  & $B$  & $-5.03622_{-0.03759}^{+0.02662}$  & $-4.74631_{-0.03380}^{+0.03171}$  & $-4.18717_{-0.02725}^{+0.02669}$\tabularnewline
 & $C$  & $-8.53154_{-0.04582}^{+0.05327}$  & $-6.43970_{-0.02674}^{+0.02594}$  & $-5.39397_{-0.01906}^{+0.01772}$\tabularnewline
\hline 
 & $A$  & $\quad0.90798_{+0.00298}^{-0.00402}$  & $\quad0.89357_{+0.00524}^{-0.00376}$  & $\quad0.86080_{+0.00425}^{-0.00328}$\tabularnewline
$q\bar{q}\to\gamma\to E^{+}E^{-}$  & $B$  & $-5.74818_{-0.02533}^{+0.03299}$  & $-5.46629_{-0.03321}^{+0.03040}$  & $-5.01876_{-0.02570}^{+0.02561}$\tabularnewline
 & $C$  & $-8.18368_{-0.05541}^{+0.04472}$  & $-6.23857_{-0.02566}^{+0.02602}$  & $-5.19925_{-0.01941}^{+0.01799}$\tabularnewline
\hline 
 & $A$  & $\quad0.94114_{+0.00760}^{-0.00508}$  & $\quad0.90624_{+0.00554}^{-0.00414}$  & $\quad0.88538_{+0.00434}^{-0.00339}$\tabularnewline
$q\bar{q}\to Z\to E^{+}E^{-}$  & $B$  & $-3.56798_{-0.04573}^{+0.03728}$  & $-3.22104_{-0.03400}^{+0.03146}$  & $-2.82716_{-0.02557}^{+0.02556}$\tabularnewline
 & $C$  & $-8.45564_{-0.03792}^{+0.04295}$  & $-6.44243_{-0.02707}^{+0.02677}$  & $-5.28945_{-0.02084}^{+0.01899}$\tabularnewline
\hline
\end{tabular}
\end{table}

%*****************************
%
%  Summary
%
%***************************** 
\section{Summary} 
We present NLO predictions for the cross sections
for new heavy quark and lepton production at three Large Hadron Collider
energies. We treat the cases of pair production of heavy quarks via
strong interactions, single heavy quark production via electroweak
interactions, and the production of exotic heavy leptons. Theoretical
uncertainties associated with the choice of the renormalization scale
and the parton distribution functions are specified. We derive a simple
and useful parametrization of our results which should assist in
experimental simulations and in phenomenological studies of new physics
models containing new heavy quarks and/or leptons.

\textbf{Acknowledgments.} E.~L.~B. is supported at Argonne by the
U.~S.\ Department of Energy under Contract No.\ DE-AC02-06CH11357.
Q.~H.~C. is supported in part by the Argonne National Laboratory
and University of Chicago Joint Theory Institute (JTI) Grant 03921-07-137,
and by the U.S.~Department of Energy under Grants No.~DE-AC02-06CH11357
and No.~DE-FG02-90ER40560. Q. H. C. is grateful to C.-P. Yuan for 
useful discussions. E.~L.~B. thanks the Galileo Galilei Institute
for Theoretical Physics in Florence for hospitality, and the Istituto
Nazionale di Fisica Nucleare (INFN) for partial support during the
final stages of this work.

{\it Note added}: While finalizing the write-up of this work, we became aware of
a recent paper~\cite{Campbell:2009gj} in which the $t$ channel production of 
single $t^\prime$ is calculated at the LHC.   Our results agree with those in 
Ref.~\cite{Campbell:2009gj}.

\appendix

%********************************************
%
%  Appendix-A: Standard model parameters
%
%******************************************** 
\section{Standard model parameters}

For our numerical evaluations, we choose the following set of SM input
parameters:\begin{eqnarray*}
& & G_{F}=1.16637\times10^{-5}{\rm GeV}^{-2}, \qquad \alpha=1/137.0359895,\\
& & m_{Z}=91.1875\,{\rm GeV}, \qquad \qquad \qquad \alpha_{s}(m_{Z})=0.1186,\\
& & m_{t}=173.1\,{\rm GeV},  \qquad \qquad \qquad \quad\ \sin^{2}\theta_{W}^{eff}=0.2314.\end{eqnarray*}
Following Ref.~\cite{Degrassi:1997iy}, we derive the $W$ boson
mass as $m_{W}=80.385\,{\rm GeV}$. Thus, the square of the weak gauge
coupling is $g^{2}=4\sqrt{2}m_{W}^{2}G_{F}$.

%******************************************************
%
%  Appendix-B: 
%
%*******************************************************  
\section{Numerical fit of the heavy quark pair production cross section
~\label{sec:smallmass}}
In this appendix we use heavy quark pair production via the QCD interaction as a sample 
illustration of how one may extend the region of applicability of a fit to a wider mass range simply by increasing the number of terms in the polynomial expression.   
We treat heavy quark production with
masses within the range 250~GeV and 700~GeV.  We choose the following 
formula to fit the cross section:
\begin{equation}
 \log\left[\frac{\sigma(m_Q,\mu)}{\rm pb}\right]
 = \frac{A}{x^2} + \frac{B}{x} + C + Dx + E x^2,     
\end{equation}
where $x=m_Q/{\rm TeV}$. 
The fitting functions at the LHC at three energies 
are given as follows:
\begin{eqnarray}
{\rm 7~TeV}  &:& 1.33969 x^2 -10.4918 x +1.36903 +1.32168/x -0.0684983/x^2,
\nonumber \\
{\rm 10~TeV} &:& 1.87135 x^2 -9.82576 x +2.34236 +1.28886/x -0.0682711/x^2,
\nonumber \\
{\rm 14~TeV} &:& 2.16753 x^2 -9.46830 x +3.33295 +1.22263/x -0.0647251/x^2,
\end{eqnarray}
where the renormalization and factorization scales are chosen as 
$\mu_{R}=\mu_{F}=\mu_0=m_Q$. Table~\ref{tab:smallmass} displays the exact
and fitted values of the NLO cross section (pb).   The numerical values of the 
exact and fitted cross sections agree very well.  

\begin{table}
\caption{The exact and fitted values of the NLO cross section (pb) for heavy
quark pair production at the LHC at three c.m. energies, where the scale
is chosen as $\mu_{R}=\mu_{F}=\mu_0=m_Q$.~\label{tab:smallmass}}
\begin{tabular}{c|cc|cc|cc}
\hline 
$m_{Q}$ & \multicolumn{2}{c|}{7TeV} & \multicolumn{2}{c|}{10TeV} & \multicolumn{2}{c}{14TeV}\tabularnewline
$(\mbox{{\rm GeV}})$ & Exact & Fit & Exact & Fit & Exact & Fit\tabularnewline
\hline 
250 & 20.51 & 20.50 & 58.34 & 58.32 & 142.10 & 142.06\tabularnewline
\hline 
300 & 7.289 & 7.291 & 22.21 & 22.21 & 57.03 & 57.04\tabularnewline
\hline 
350 & 2.940 & 2.939 & 14.38 & 14.38 & 25.78 & 25.78\tabularnewline
\hline 
400 & 1.301 & 1.301 & 9.565 & 9.561 & 12.74 & 12.74\tabularnewline
\hline 
450 & 0.617 & 0.618 & 2.285 & 2.286 & 6.742 & 6.743\tabularnewline
\hline 
500 & 0.310 & 0.310 & 1.224 & 1.224 & 3.771 & 3.770\tabularnewline
\hline 
550 & 0.162 & 0.162 & 0.685 & 0.685 & 2.203 & 2.204\tabularnewline
\hline 
600 & 0.088 & 0.088 & 0.398 & 0.398 & 1.336 & 1.337\tabularnewline
\hline 
650 & 0.049 & 0.049 & 0.239 & 0.239 & 0.837 & 0.837\tabularnewline
\hline 
700 & 0.028 & 0.028 & 0.147 & 0.147 & 0.539 & 0.539\tabularnewline
\hline
\end{tabular}

\end{table}

%******************************************************
%
%  Appendix-C: Influence of parton luminosities on
%              the heavy-fermion mass dependence
%
%*******************************************************  
\section{Influence of parton luminosities on the heavy fermion mass dependence~\label{sec:pdf}}

We comment in this appendix on the origins of the remarkably simple 
analytic expression Eq.~\ref{eq:formula} and on its range of applicability 
in the heavy fermion mass.  The expression works well for 
strong pair production of heavy quarks, as well 
as for electroweak single heavy fermion production and electroweak pair 
production of heavy fermions.  This apparent process independence suggests 
that the heavy fermion mass dependence of the parton 
luminosities~\cite{Eichten:1984eu,ESW:book,Quigg:2009gg} is playing a dominant role.   

The quark-antiquark and gluon-gluon contributions to the cross section for 
heavy quark pair production at hadron colliders can be written as 
\begin{eqnarray}
\sigma_{q \bar{q}}(s) & = & \sum_{q=u,d}\int_{\tau_{min}}^{1}d\tau\int_{\tau}^{1}\frac{dx_{a}}{x_{a}}\left[f_{q/P}\left(x_{a},\mu_{0}\right)f_{\bar{q}/P}\left(\frac{\tau}{x_{a}},\mu_{0}\right)+\left(u\leftrightarrow\bar{u}\right)\right]\frac{1}{\hat{s}}\left[\hat{s}\hat{\sigma}_{q\bar{q}\to Q\bar{Q}}(\hat{s})\right],\nonumber \\
\sigma_{gg}(s) & = & \int_{\tau_{min}}^{1}d\tau\int_{\tau}^{1}\frac{dx_{a}}{x_{a}}f_{g/P}\left(x_{a},\mu_{0}\right)f_{g/P}\left(\frac{\tau}{x_{a}},\mu_{0}\right)\frac{1}{\hat{s}}\left[\hat{s}\hat{\sigma}_{q\bar{q}\to Q\bar{Q}}(\hat{s})\right],\label{eq:s}
\end{eqnarray}
where $\hat{\sigma}$ is the operative parton-level cross section,
$\tau_{min}=4m_{Q}^{2}/s$, and $\hat{s}=\tau s$ with $s$ being the square of the 
c.m. energy of LHC.  In order to represent a process independent situation, we
approximate the (dimensionless) square brackets by one. As a result,
the only remaining effect of the heavy quarks appears in the lower limit
of integration $\tau_{min}$.   We choose $\mu_{R}=\mu_{F}=m_{Q}$
and use the CTEQ6.6M average PDF set.

\begin{figure}
\includegraphics[scale=0.6]{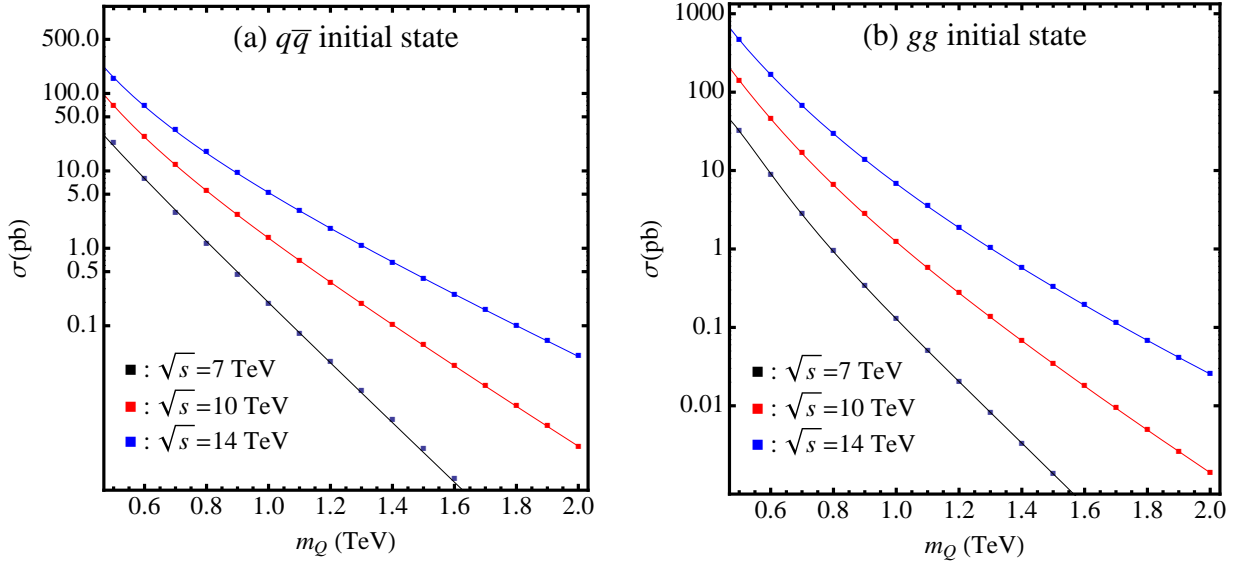}
\caption{Fit to the trial scattering cross sections in the range of 
$0.5 < m_Q < 2.0$~TeV: (a) for the quark-antiquark annihilation channel,
and (b) for the gluon fusion channel. \label{fig:fake-well}}
\end{figure}

Using Eq.~\ref{eq:s},  we generate ``cross sections'' as a function of 
$m_Q$, and we then fit the results with Eq.~\ref{eq:formula}.   The 
agreement is excellent for the $q \bar{q}$ and $gg$ cases over the range 
of heavy quark masses shown in Fig.~\ref{fig:fake-well}, $0.5 < m_Q < 2.0$~TeV.
The square symbols in the figure are the predicted cross sections,  
and the solid curves represents the fits.   

\begin{figure}
\includegraphics[scale=0.6]{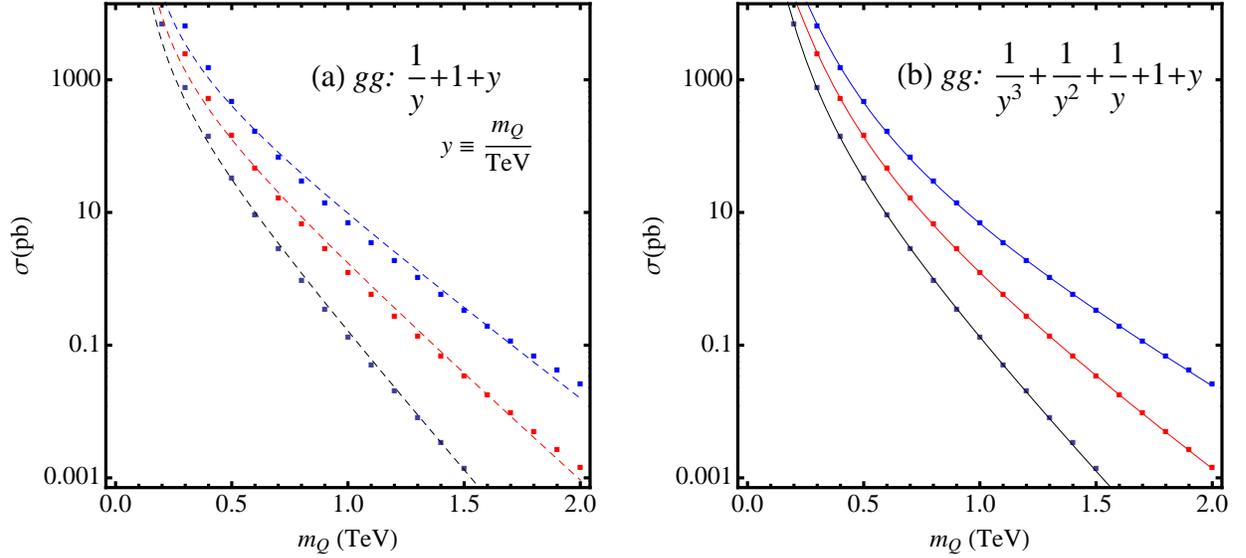}
\caption{Fit to the trial scattering cross section for the $gg$ channel extended 
into the region of small $m_Q$.\label{fig:Fake_bad} }
\end{figure}

If we try to extend our study to smaller masses than shown in 
Fig.~\ref{fig:fake-well} we find that the simple expression 
Eq.~\ref{eq:formula} is no longer adequate, as shown in 
Fig.~\ref{fig:Fake_bad}(a).   When $\tau_{min}$ is too small, 
we sample parton densities in the region of small Bjorken $x$ where
they grow rapidly with decreasing $x$.   More terms in
the polynomial are needed for a better fit.   We can
fit the gluon fusion cross section well with the addition of two 
more terms; see Fig.~\ref{fig:Fake_bad}(b)

\newpage
\bibliography{reference}

\begin{thebibliography}{42}
\expandafter\ifx\csname natexlab\endcsname\relax\def\natexlab#1{#1}\fi
\expandafter\ifx\csname bibnamefont\endcsname\relax
  \def\bibnamefont#1{#1}\fi
\expandafter\ifx\csname bibfnamefont\endcsname\relax
  \def\bibfnamefont#1{#1}\fi
\expandafter\ifx\csname citenamefont\endcsname\relax
  \def\citenamefont#1{#1}\fi
\expandafter\ifx\csname url\endcsname\relax
  \def\url#1{\texttt{#1}}\fi
\expandafter\ifx\csname urlprefix\endcsname\relax\def\urlprefix{URL }\fi
\providecommand{\bibinfo}[2]{#2}
\providecommand{\eprint}[2][]{\url{#2}}

\bibitem[{\citenamefont{Alwall et~al.}(2007)}]{Alwall:2006bx}
\bibinfo{author}{\bibfnamefont{J.}~\bibnamefont{Alwall}} \bibnamefont{et~al.},
  \bibinfo{journal}{Eur. Phys. J.} \textbf{\bibinfo{volume}{C49}},
  \bibinfo{pages}{791} (\bibinfo{year}{2007}), \eprint{hep-ph/0607115}.

\bibitem[{\citenamefont{Kribs et~al.}(2007)\citenamefont{Kribs, Plehn,
  Spannowsky, and Tait}}]{Kribs:2007nz}
\bibinfo{author}{\bibfnamefont{G.~D.} \bibnamefont{Kribs}},
  \bibinfo{author}{\bibfnamefont{T.}~\bibnamefont{Plehn}},
  \bibinfo{author}{\bibfnamefont{M.}~\bibnamefont{Spannowsky}},
  \bibnamefont{and} \bibinfo{author}{\bibfnamefont{T.~M.~P.}
  \bibnamefont{Tait}}, \bibinfo{journal}{Phys. Rev.}
  \textbf{\bibinfo{volume}{D76}}, \bibinfo{pages}{075016}
  (\bibinfo{year}{2007}), \eprint{0706.3718}.

\bibitem[{\citenamefont{Soni et~al.}(2008)\citenamefont{Soni, Alok, Giri,
  Mohanta, and Nandi}}]{Soni:2008bc}
\bibinfo{author}{\bibfnamefont{A.}~\bibnamefont{Soni}},
  \bibinfo{author}{\bibfnamefont{A.~K.} \bibnamefont{Alok}},
  \bibinfo{author}{\bibfnamefont{A.}~\bibnamefont{Giri}},
  \bibinfo{author}{\bibfnamefont{R.}~\bibnamefont{Mohanta}}, \bibnamefont{and}
  \bibinfo{author}{\bibfnamefont{S.}~\bibnamefont{Nandi}}
  (\bibinfo{year}{2008}), \eprint{0807.1971}.

\bibitem[{\citenamefont{Bobrowski et~al.}(2009)\citenamefont{Bobrowski, Lenz,
  Riedl, and Rohrwild}}]{Bobrowski:2009ng}
\bibinfo{author}{\bibfnamefont{M.}~\bibnamefont{Bobrowski}},
  \bibinfo{author}{\bibfnamefont{A.}~\bibnamefont{Lenz}},
  \bibinfo{author}{\bibfnamefont{J.}~\bibnamefont{Riedl}}, \bibnamefont{and}
  \bibinfo{author}{\bibfnamefont{J.}~\bibnamefont{Rohrwild}},
  \bibinfo{journal}{Phys. Rev.} \textbf{\bibinfo{volume}{D79}},
  \bibinfo{pages}{113006} (\bibinfo{year}{2009}), \eprint{0902.4883}.

\bibitem[{\citenamefont{Chanowitz}(2009)}]{Chanowitz:2009mz}
\bibinfo{author}{\bibfnamefont{M.~S.} \bibnamefont{Chanowitz}},
  \bibinfo{journal}{Phys. Rev.} \textbf{\bibinfo{volume}{D79}},
  \bibinfo{pages}{113008} (\bibinfo{year}{2009}), \eprint{0904.3570}.

\bibitem[{\citenamefont{Holdom et~al.}(2009)\citenamefont{Holdom, Hou, Hurth,
  Mangano, Sultansoy, and Unel}}]{Holdom:2009rf}
\bibinfo{author}{\bibfnamefont{B.}~\bibnamefont{Holdom}},
  \bibinfo{author}{\bibfnamefont{W.~S.} \bibnamefont{Hou}},
  \bibinfo{author}{\bibfnamefont{T.}~\bibnamefont{Hurth}},
  \bibinfo{author}{\bibfnamefont{M.~L.} \bibnamefont{Mangano}},
  \bibinfo{author}{\bibfnamefont{S.}~\bibnamefont{Sultansoy}},
  \bibnamefont{and} \bibinfo{author}{\bibfnamefont{G.}~\bibnamefont{Unel}}
  (\bibinfo{year}{2009}), \eprint{0904.4698}.

\bibitem[{\citenamefont{Soni}(2009)}]{Soni:2009fg}
\bibinfo{author}{\bibfnamefont{A.}~\bibnamefont{Soni}} (\bibinfo{year}{2009}),
  \eprint{0907.2057}.

\bibitem[{\citenamefont{Eilam et~al.}(2009)\citenamefont{Eilam, Melic, and
  Trampetic}}]{Eilam:2009hz}
\bibinfo{author}{\bibfnamefont{G.}~\bibnamefont{Eilam}},
  \bibinfo{author}{\bibfnamefont{B.}~\bibnamefont{Melic}}, \bibnamefont{and}
  \bibinfo{author}{\bibfnamefont{J.}~\bibnamefont{Trampetic}}
  (\bibinfo{year}{2009}), \eprint{0909.3227}.

\bibitem[{\citenamefont{Amsler et~al.}(2008)}]{Amsler:2008zzb}
\bibinfo{author}{\bibfnamefont{C.}~\bibnamefont{Amsler}} \bibnamefont{et~al.}
  (\bibinfo{collaboration}{Particle Data Group}), \bibinfo{journal}{Phys.
  Lett.} \textbf{\bibinfo{volume}{B667}}, \bibinfo{pages}{1}
  (\bibinfo{year}{2008}).

\bibitem[{\citenamefont{del Aguila et~al.}(2000)\citenamefont{del Aguila,
  Perez-Victoria, and Santiago}}]{delAguila:2000rc}
\bibinfo{author}{\bibfnamefont{F.}~\bibnamefont{del Aguila}},
  \bibinfo{author}{\bibfnamefont{M.}~\bibnamefont{Perez-Victoria}},
  \bibnamefont{and} \bibinfo{author}{\bibfnamefont{J.}~\bibnamefont{Santiago}},
  \bibinfo{journal}{JHEP} \textbf{\bibinfo{volume}{09}}, \bibinfo{pages}{011}
  (\bibinfo{year}{2000}), \eprint{hep-ph/0007316}.

\bibitem[{\citenamefont{Dobrescu et~al.}(2009)\citenamefont{Dobrescu, Kong, and
  Mahbubani}}]{Dobrescu:2009vz}
\bibinfo{author}{\bibfnamefont{B.~A.} \bibnamefont{Dobrescu}},
  \bibinfo{author}{\bibfnamefont{K.}~\bibnamefont{Kong}}, \bibnamefont{and}
  \bibinfo{author}{\bibfnamefont{R.}~\bibnamefont{Mahbubani}},
  \bibinfo{journal}{JHEP} \textbf{\bibinfo{volume}{06}}, \bibinfo{pages}{001}
  (\bibinfo{year}{2009}), \eprint{0902.0792}.

\bibitem[{\citenamefont{Choudhury et~al.}(2002)\citenamefont{Choudhury, Tait,
  and Wagner}}]{Choudhury:2001hs}
\bibinfo{author}{\bibfnamefont{D.}~\bibnamefont{Choudhury}},
  \bibinfo{author}{\bibfnamefont{T.~M.~P.} \bibnamefont{Tait}},
  \bibnamefont{and} \bibinfo{author}{\bibfnamefont{C.~E.~M.}
  \bibnamefont{Wagner}}, \bibinfo{journal}{Phys. Rev.}
  \textbf{\bibinfo{volume}{D65}}, \bibinfo{pages}{053002}
  (\bibinfo{year}{2002}), \eprint{hep-ph/0109097}.

\bibitem[{\citenamefont{Arkani-Hamed et~al.}(2002)\citenamefont{Arkani-Hamed,
  Cohen, Katz, and Nelson}}]{ArkaniHamed:2002qy}
\bibinfo{author}{\bibfnamefont{N.}~\bibnamefont{Arkani-Hamed}},
  \bibinfo{author}{\bibfnamefont{A.~G.} \bibnamefont{Cohen}},
  \bibinfo{author}{\bibfnamefont{E.}~\bibnamefont{Katz}}, \bibnamefont{and}
  \bibinfo{author}{\bibfnamefont{A.~E.} \bibnamefont{Nelson}},
  \bibinfo{journal}{JHEP} \textbf{\bibinfo{volume}{07}}, \bibinfo{pages}{034}
  (\bibinfo{year}{2002}), \eprint{hep-ph/0206021}.

\bibitem[{\citenamefont{Perelstein et~al.}(2004)\citenamefont{Perelstein,
  Peskin, and Pierce}}]{Perelstein:2003wd}
\bibinfo{author}{\bibfnamefont{M.}~\bibnamefont{Perelstein}},
  \bibinfo{author}{\bibfnamefont{M.~E.} \bibnamefont{Peskin}},
  \bibnamefont{and} \bibinfo{author}{\bibfnamefont{A.}~\bibnamefont{Pierce}},
  \bibinfo{journal}{Phys. Rev.} \textbf{\bibinfo{volume}{D69}},
  \bibinfo{pages}{075002} (\bibinfo{year}{2004}), \eprint{hep-ph/0310039}.

\bibitem[{\citenamefont{Han et~al.}(2003)\citenamefont{Han, Logan, McElrath,
  and Wang}}]{Han:2003wu}
\bibinfo{author}{\bibfnamefont{T.}~\bibnamefont{Han}},
  \bibinfo{author}{\bibfnamefont{H.~E.} \bibnamefont{Logan}},
  \bibinfo{author}{\bibfnamefont{B.}~\bibnamefont{McElrath}}, \bibnamefont{and}
  \bibinfo{author}{\bibfnamefont{L.-T.} \bibnamefont{Wang}},
  \bibinfo{journal}{Phys. Rev.} \textbf{\bibinfo{volume}{D67}},
  \bibinfo{pages}{095004} (\bibinfo{year}{2003}), \eprint{hep-ph/0301040}.

\bibitem[{\citenamefont{Cheng et~al.}(2006)\citenamefont{Cheng, Low, and
  Wang}}]{Cheng:2005as}
\bibinfo{author}{\bibfnamefont{H.-C.} \bibnamefont{Cheng}},
  \bibinfo{author}{\bibfnamefont{I.}~\bibnamefont{Low}}, \bibnamefont{and}
  \bibinfo{author}{\bibfnamefont{L.-T.} \bibnamefont{Wang}},
  \bibinfo{journal}{Phys. Rev.} \textbf{\bibinfo{volume}{D74}},
  \bibinfo{pages}{055001} (\bibinfo{year}{2006}), \eprint{hep-ph/0510225}.

\bibitem[{\citenamefont{del Aguila et~al.}(1990)\citenamefont{del Aguila,
  Ametller, Kane, and Vidal}}]{delAguila:1989rq}
\bibinfo{author}{\bibfnamefont{F.}~\bibnamefont{del Aguila}},
  \bibinfo{author}{\bibfnamefont{L.}~\bibnamefont{Ametller}},
  \bibinfo{author}{\bibfnamefont{G.~L.} \bibnamefont{Kane}}, \bibnamefont{and}
  \bibinfo{author}{\bibfnamefont{J.}~\bibnamefont{Vidal}},
  \bibinfo{journal}{Nucl. Phys.} \textbf{\bibinfo{volume}{B334}},
  \bibinfo{pages}{1} (\bibinfo{year}{1990}).

\bibitem[{\citenamefont{Aguilar-Saavedra}(2005)}]{AguilarSaavedra:2005pv}
\bibinfo{author}{\bibfnamefont{J.~A.} \bibnamefont{Aguilar-Saavedra}},
  \bibinfo{journal}{Phys. Lett.} \textbf{\bibinfo{volume}{B625}},
  \bibinfo{pages}{234} (\bibinfo{year}{2005}), \eprint{hep-ph/0506187}.

\bibitem[{\citenamefont{Matsumoto et~al.}(2007)\citenamefont{Matsumoto, Nojiri,
  and Nomura}}]{Matsumoto:2006ws}
\bibinfo{author}{\bibfnamefont{S.}~\bibnamefont{Matsumoto}},
  \bibinfo{author}{\bibfnamefont{M.~M.} \bibnamefont{Nojiri}},
  \bibnamefont{and} \bibinfo{author}{\bibfnamefont{D.}~\bibnamefont{Nomura}},
  \bibinfo{journal}{Phys. Rev.} \textbf{\bibinfo{volume}{D75}},
  \bibinfo{pages}{055006} (\bibinfo{year}{2007}), \eprint{hep-ph/0612249}.

\bibitem[{\citenamefont{Freitas and Wyler}(2006)}]{Freitas:2006vy}
\bibinfo{author}{\bibfnamefont{A.}~\bibnamefont{Freitas}} \bibnamefont{and}
  \bibinfo{author}{\bibfnamefont{D.}~\bibnamefont{Wyler}},
  \bibinfo{journal}{JHEP} \textbf{\bibinfo{volume}{11}}, \bibinfo{pages}{061}
  (\bibinfo{year}{2006}), \eprint{hep-ph/0609103}.

\bibitem[{\citenamefont{Meade and Reece}(2006)}]{Meade:2006dw}
\bibinfo{author}{\bibfnamefont{P.}~\bibnamefont{Meade}} \bibnamefont{and}
  \bibinfo{author}{\bibfnamefont{M.}~\bibnamefont{Reece}},
  \bibinfo{journal}{Phys. Rev.} \textbf{\bibinfo{volume}{D74}},
  \bibinfo{pages}{015010} (\bibinfo{year}{2006}), \eprint{hep-ph/0601124}.

\bibitem[{\citenamefont{Cao et~al.}(2008)\citenamefont{Cao, Li, and
  Yuan}}]{Cao:2006wk}
\bibinfo{author}{\bibfnamefont{Q.-H.} \bibnamefont{Cao}},
  \bibinfo{author}{\bibfnamefont{C.~S.} \bibnamefont{Li}}, \bibnamefont{and}
  \bibinfo{author}{\bibfnamefont{C.~P.} \bibnamefont{Yuan}},
  \bibinfo{journal}{Phys. Lett.} \textbf{\bibinfo{volume}{B668}},
  \bibinfo{pages}{24} (\bibinfo{year}{2008}), \eprint{hep-ph/0612243}.

\bibitem[{\citenamefont{Belyaev et~al.}(2006)\citenamefont{Belyaev, Chen, Tobe,
  and Yuan}}]{Belyaev:2006jh}
\bibinfo{author}{\bibfnamefont{A.}~\bibnamefont{Belyaev}},
  \bibinfo{author}{\bibfnamefont{C.-R.} \bibnamefont{Chen}},
  \bibinfo{author}{\bibfnamefont{K.}~\bibnamefont{Tobe}}, \bibnamefont{and}
  \bibinfo{author}{\bibfnamefont{C.~P.} \bibnamefont{Yuan}},
  \bibinfo{journal}{Phys. Rev.} \textbf{\bibinfo{volume}{D74}},
  \bibinfo{pages}{115020} (\bibinfo{year}{2006}), \eprint{hep-ph/0609179}.

\bibitem[{\citenamefont{Contino and Servant}(2008)}]{Contino:2008hi}
\bibinfo{author}{\bibfnamefont{R.}~\bibnamefont{Contino}} \bibnamefont{and}
  \bibinfo{author}{\bibfnamefont{G.}~\bibnamefont{Servant}},
  \bibinfo{journal}{JHEP} \textbf{\bibinfo{volume}{06}}, \bibinfo{pages}{026}
  (\bibinfo{year}{2008}), \eprint{0801.1679}.

\bibitem[{\citenamefont{Aguilar-Saavedra}(2009{\natexlab{a}})}]{AguilarSaavedr%
a:2009es}
\bibinfo{author}{\bibfnamefont{J.~A.} \bibnamefont{Aguilar-Saavedra}}
  (\bibinfo{year}{2009}{\natexlab{a}}), \eprint{0907.3155}.

\bibitem[{\citenamefont{Nason et~al.}(1988)\citenamefont{Nason, Dawson, and
  Ellis}}]{Nason:1987xz}
\bibinfo{author}{\bibfnamefont{P.}~\bibnamefont{Nason}},
  \bibinfo{author}{\bibfnamefont{S.}~\bibnamefont{Dawson}}, \bibnamefont{and}
  \bibinfo{author}{\bibfnamefont{R.~K.} \bibnamefont{Ellis}},
  \bibinfo{journal}{Nucl. Phys.} \textbf{\bibinfo{volume}{B303}},
  \bibinfo{pages}{607} (\bibinfo{year}{1988}).

\bibitem[{\citenamefont{Beenakker et~al.}(1989)\citenamefont{Beenakker, Kuijf,
  van Neerven, and Smith}}]{Beenakker:1988bq}
\bibinfo{author}{\bibfnamefont{W.}~\bibnamefont{Beenakker}},
  \bibinfo{author}{\bibfnamefont{H.}~\bibnamefont{Kuijf}},
  \bibinfo{author}{\bibfnamefont{W.~L.} \bibnamefont{van Neerven}},
  \bibnamefont{and} \bibinfo{author}{\bibfnamefont{J.}~\bibnamefont{Smith}},
  \bibinfo{journal}{Phys. Rev.} \textbf{\bibinfo{volume}{D40}},
  \bibinfo{pages}{54} (\bibinfo{year}{1989}).

\bibitem[{\citenamefont{Nason et~al.}(1989)\citenamefont{Nason, Dawson, and
  Ellis}}]{Nason:1989zy}
\bibinfo{author}{\bibfnamefont{P.}~\bibnamefont{Nason}},
  \bibinfo{author}{\bibfnamefont{S.}~\bibnamefont{Dawson}}, \bibnamefont{and}
  \bibinfo{author}{\bibfnamefont{R.~K.} \bibnamefont{Ellis}},
  \bibinfo{journal}{Nucl. Phys.} \textbf{\bibinfo{volume}{B327}},
  \bibinfo{pages}{49} (\bibinfo{year}{1989}).

\bibitem[{\citenamefont{Beenakker et~al.}(1991)\citenamefont{Beenakker, van
  Neerven, Meng, Schuler, and Smith}}]{Beenakker:1990maa}
\bibinfo{author}{\bibfnamefont{W.}~\bibnamefont{Beenakker}},
  \bibinfo{author}{\bibfnamefont{W.~L.} \bibnamefont{van Neerven}},
  \bibinfo{author}{\bibfnamefont{R.}~\bibnamefont{Meng}},
  \bibinfo{author}{\bibfnamefont{G.~A.} \bibnamefont{Schuler}},
  \bibnamefont{and} \bibinfo{author}{\bibfnamefont{J.}~\bibnamefont{Smith}},
  \bibinfo{journal}{Nucl. Phys.} \textbf{\bibinfo{volume}{B351}},
  \bibinfo{pages}{507} (\bibinfo{year}{1991}).

\bibitem[{\citenamefont{Stump et~al.}(2003)}]{Stump:2003yu}
\bibinfo{author}{\bibfnamefont{D.}~\bibnamefont{Stump}} \bibnamefont{et~al.},
  \bibinfo{journal}{JHEP} \textbf{\bibinfo{volume}{10}}, \bibinfo{pages}{046}
  (\bibinfo{year}{2003}), \eprint{hep-ph/0303013}.

\bibitem[{\citenamefont{Martin et~al.}(2009)\citenamefont{Martin, Stirling,
  Thorne, and Watt}}]{Martin:2009iq}
\bibinfo{author}{\bibfnamefont{A.~D.} \bibnamefont{Martin}},
  \bibinfo{author}{\bibfnamefont{W.~J.} \bibnamefont{Stirling}},
  \bibinfo{author}{\bibfnamefont{R.~S.} \bibnamefont{Thorne}},
  \bibnamefont{and} \bibinfo{author}{\bibfnamefont{G.}~\bibnamefont{Watt}},
  \bibinfo{journal}{Eur. Phys. J.} \textbf{\bibinfo{volume}{C63}},
  \bibinfo{pages}{189} (\bibinfo{year}{2009}), \eprint{0901.0002}.

\bibitem[{\citenamefont{Ball et~al.}(2009)}]{Ball:2008by}
\bibinfo{author}{\bibfnamefont{R.~D.} \bibnamefont{Ball}} \bibnamefont{et~al.}
  (\bibinfo{collaboration}{NNPDF}), \bibinfo{journal}{Nucl. Phys.}
  \textbf{\bibinfo{volume}{B809}}, \bibinfo{pages}{1} (\bibinfo{year}{2009}),
  \eprint{0808.1231}.

\bibitem[{\citenamefont{Kidonakis and Vogt}(2008)}]{Kidonakis:2008mu}
\bibinfo{author}{\bibfnamefont{N.}~\bibnamefont{Kidonakis}} \bibnamefont{and}
  \bibinfo{author}{\bibfnamefont{R.}~\bibnamefont{Vogt}},
  \bibinfo{journal}{Phys. Rev.} \textbf{\bibinfo{volume}{D78}},
  \bibinfo{pages}{074005} (\bibinfo{year}{2008}), \eprint{0805.3844}.

\bibitem[{\citenamefont{Stelzer et~al.}(1997)\citenamefont{Stelzer, Sullivan,
  and Willenbrock}}]{Stelzer:1997ns}
\bibinfo{author}{\bibfnamefont{T.}~\bibnamefont{Stelzer}},
  \bibinfo{author}{\bibfnamefont{Z.}~\bibnamefont{Sullivan}}, \bibnamefont{and}
  \bibinfo{author}{\bibfnamefont{S.}~\bibnamefont{Willenbrock}},
  \bibinfo{journal}{Phys. Rev.} \textbf{\bibinfo{volume}{D56}},
  \bibinfo{pages}{5919} (\bibinfo{year}{1997}), \eprint{hep-ph/9705398}.

\bibitem[{\citenamefont{del Aguila and
  Aguilar-Saavedra}(2009{\natexlab{a}})}]{delAguila:2008cj}
\bibinfo{author}{\bibfnamefont{F.}~\bibnamefont{del Aguila}} \bibnamefont{and}
  \bibinfo{author}{\bibfnamefont{J.~A.} \bibnamefont{Aguilar-Saavedra}},
  \bibinfo{journal}{Nucl. Phys.} \textbf{\bibinfo{volume}{B813}},
  \bibinfo{pages}{22} (\bibinfo{year}{2009}{\natexlab{a}}), \eprint{0808.2468}.

\bibitem[{\citenamefont{Aguilar-Saavedra}(2009{\natexlab{b}})}]{AguilarSaavedr%
a:2009ik}
\bibinfo{author}{\bibfnamefont{J.~A.} \bibnamefont{Aguilar-Saavedra}}
  (\bibinfo{year}{2009}{\natexlab{b}}), \eprint{0905.2221}.

\bibitem[{\citenamefont{del Aguila and
  Aguilar-Saavedra}(2009{\natexlab{b}})}]{delAguila:2008hw}
\bibinfo{author}{\bibfnamefont{F.}~\bibnamefont{del Aguila}} \bibnamefont{and}
  \bibinfo{author}{\bibfnamefont{J.~A.} \bibnamefont{Aguilar-Saavedra}},
  \bibinfo{journal}{Phys. Lett.} \textbf{\bibinfo{volume}{B672}},
  \bibinfo{pages}{158} (\bibinfo{year}{2009}{\natexlab{b}}),
  \eprint{0809.2096}.

\bibitem[{\citenamefont{Campbell et~al.}(2009)\citenamefont{Campbell, Frederix,
  Maltoni, and Tramontano}}]{Campbell:2009gj}
\bibinfo{author}{\bibfnamefont{J.~M.} \bibnamefont{Campbell}},
  \bibinfo{author}{\bibfnamefont{R.}~\bibnamefont{Frederix}},
  \bibinfo{author}{\bibfnamefont{F.}~\bibnamefont{Maltoni}}, \bibnamefont{and}
  \bibinfo{author}{\bibfnamefont{F.}~\bibnamefont{Tramontano}}
  (\bibinfo{year}{2009}), \eprint{0907.3933}.

\bibitem[{\citenamefont{Degrassi et~al.}(1998)\citenamefont{Degrassi, Gambino,
  Passera, and Sirlin}}]{Degrassi:1997iy}
\bibinfo{author}{\bibfnamefont{G.}~\bibnamefont{Degrassi}},
  \bibinfo{author}{\bibfnamefont{P.}~\bibnamefont{Gambino}},
  \bibinfo{author}{\bibfnamefont{M.}~\bibnamefont{Passera}}, \bibnamefont{and}
  \bibinfo{author}{\bibfnamefont{A.}~\bibnamefont{Sirlin}},
  \bibinfo{journal}{Phys. Lett.} \textbf{\bibinfo{volume}{B418}},
  \bibinfo{pages}{209} (\bibinfo{year}{1998}), \eprint{hep-ph/9708311}.

\bibitem[{\citenamefont{Eichten et~al.}(1984)\citenamefont{Eichten, Hinchliffe,
  Lane, and Quigg}}]{Eichten:1984eu}
\bibinfo{author}{\bibfnamefont{E.}~\bibnamefont{Eichten}},
  \bibinfo{author}{\bibfnamefont{I.}~\bibnamefont{Hinchliffe}},
  \bibinfo{author}{\bibfnamefont{K.~D.} \bibnamefont{Lane}}, \bibnamefont{and}
  \bibinfo{author}{\bibfnamefont{C.}~\bibnamefont{Quigg}},
  \bibinfo{journal}{Rev. Mod. Phys.} \textbf{\bibinfo{volume}{56}},
  \bibinfo{pages}{579} (\bibinfo{year}{1984}).

\bibitem[{\citenamefont{Ellis et~al.}(2003)\citenamefont{Ellis, Stirling, and
  Webber}}]{ESW:book}
\bibinfo{author}{\bibfnamefont{R.~K.} \bibnamefont{Ellis}},
  \bibinfo{author}{\bibfnamefont{W.~J.} \bibnamefont{Stirling}},
  \bibnamefont{and} \bibinfo{author}{\bibfnamefont{B.~R.}
  \bibnamefont{Webber}}, \emph{\bibinfo{title}{QCD and Collider Physics
  (Cambridge Monographs on Particle Physics, Nuclear Physics and Cosmology)}}
  (\bibinfo{publisher}{Cambridge University Press}, \bibinfo{year}{2003}).

\bibitem[{\citenamefont{Quigg}(2009)}]{Quigg:2009gg}
\bibinfo{author}{\bibfnamefont{C.}~\bibnamefont{Quigg}} (\bibinfo{year}{2009}),
  \eprint{0908.3660}.

\end{thebibliography}

\end{document}